# Physical factors governing the shape of the Miram curve knee in thermionic emission

Dongzheng Chen, Ryan Jacobs, Dane Morgan, and John Booske, *Fellow, IEEE*

*Abstract*—In a current density versus temperature (*J-T*) (Miram) curve in thermionic electron emission, experimental measurements demonstrate there is a smooth transition between the exponential region and the saturated emission regions, which is sometimes referred to as the "roll-off" or "Miram curve knee". The shape of the Miram curve knee is an important figure of merit for thermionic vacuum cathodes. Specifically, cathodes with a sharp Miram curve knee at low temperature with a flat saturated emission current are typically preferred. Our previous work on modeling nonuniform thermionic emission revealed that the space charge effect and patch field effect are key pieces of physics which impact the shape of the Miram curve knee. This work provides a more complete understanding of the physical factors connecting these physical effects and their relative impact on the shape of the knee, including the smoothness, the temperature, and the flatness of the saturated emission current density. For our analyses, we use a periodic, equal-width striped ("zebra crossing") work function distribution as a model system and illustrate how the space charge and patch field effects restrict the emission current density near the Miram curve knee. The results indicate there are three main physical parameters which significantly impact the shape of the Miram curve. Such physical knowledge directly connects the patch size, work function values, anode-cathode voltage, and anode-cathode gap distance to the shape of the Miram curve, providing new understanding and a guide to the design of thermionic cathodes used as electron sources in vacuum electronic devices (VEDs).

## I. Introduction

THERMIONIC cathodes provide the electron source in numerous vacuum electronic devices (VEDs) used in civilian, industrial, and scientific applications, such as communication devices, ion thrusters, thermionic energy converters, and free electron lasers.[1]–[3] The total emission current density divided by the cathode area, referred to as the cathode-averaged emission current density, $J$, is a key property in thermionic emission. $J$ is affected by many parameters including temperature $T$, anode-cathode voltage $V_{AK}$, diode geometry, etc. $J - T$ curves normalized by the extrapolated full-space-charge-limited emission current density $J_{FSCL}$ are referred to as Miram curves, named after George Miram.[4] Miram (or $J - T$) curves are widely used as a figure of merit to evaluate the cathode performance. Experimental observations show there is a smooth transition, called the "roll off" or Miram curve knee, between the exponentially growing (temperature-limited or TL) region and the saturated emission current (full-space-charge-limited or FSCL) region. The shape characterizing the Miram curve knee is important as thermionic cathodes are almost always operated at a temperature somewhat above the Miram curve knee. For VEDs used in space applications, the operating point may be slightly above the knee temperature to realize stable cathode operation at the lowest temperature, thus maximizing cathode lifetime and reducing the input power required to keep the cathode at high temperature. For non-space applications, a higher temperature may be used in order to maintain space charge limited emission and mitigate potential fluctuations in the emitted current density that may occur if the cathode enters the temperature-limited emission region.[5], [6] Thus, knowledge of the shape of the Miram curve knee region, including the smoothness, the temperature, and the flatness of the saturated emission current density, can inform desired operational parameters of VEDs using thermionic cathodes.

Our recent work[7] developed a physics-based model of nonuniform thermionic emission that includes the effects of space charge, patch fields, and Schottky barrier lowering, giving a mathematical method to calculate the emission current from a cathode with a spatially heterogeneous work function distribution in a parallel diode. This model predicts a smooth Miram curve knee for a model cathode surface consisting of a 2-D checkerboard work function distribution. The results of this model reveal that the smoothness of the Miram curve knee arises as a natural consequence of the physics of nonuniform thermionic emission, and that the space charge and patch field effects have more significant impact on smoothing the Miram curve knee than the Schottky barrier lowering. Further study[8] showed that when a 2-D work function distribution obtained

This work was funded by the Defense Advanced Research Projects Agency (DARPA) under grant reference number N66001-16-1-4043 through the Innovative Vacuum Electronic Science and Technology (INVEST) program.

Dongzheng Chen, Ryan Jacobs, and Dane Morgan are with the Department of Materials Science and Engineering, University of Wisconsin-Madison, Madison, WI 53706, USA.

John Booske is with the Electrical and Computer Engineering Department, University of Wisconsin–Madison, Madison, WI 53706, USA (e-mail: jhbooske@wisc.edu).

from a commercial tungsten dispenser cathode is used with this nonuniform emission model, the model is able to predict $J-T$ and $J-V$ emission curves in near quantitative agreement with experimental data, and accurately predicts the shape of the Miram curve knee.

Our previously published work[7] on the nonuniform emission model successfully included the predictions of key characteristics of the Miram curve, including the smooth knee, but it did not include a detailed examination of the different physical factors connecting the effects of space charge and patch fields and their relative impact on the shape of the Miram curve knee. In this work, we use our nonuniform emission model to further study the effects of space charge and patch fields on the shape of the Miram curve knee, with the goal of understanding which physical parameters have the most significant impact on the shape of the knee, and how the typical ranges of the values of these parameters impact the knee. Knowledge of how the surface microstructure, including the work function values and the emitting patch area sizes affect the shape of the Miram curve knee would provide new understanding of cathode materials design and engineering. Similarly, incorporating the effects of diode geometry and anode-cathode voltage could improve the design of electron gun fixtures used in VEDs. This work also provides new insights in the analysis of experimental Miram curves, which are useful for developing and understanding the performance of novel thermionic cathodes using new materials.

## II. METHODS AND THEORY

### A. Zebra crossing work function distribution

Checkerboard work function distributions[7], [9]–[14] and equal-width periodic stripes (or "zebra crossing") work function distributions[5], [13], [15], [16] are two model work function distributions used in previous studies of nonuniform thermionic emission. As illustrated in Ref. [13], many of the fundamental properties of electron emission from the two-dimensional checkerboard distribution are observable with the simpler, one-dimensional, zebra-crossing distribution. Therefore, in this work we use the zebra crossing surface model to analyze the effects of space charge and patch fields on the shape of Miram curve.

Fig. 1 illustrates the zebra crossing model of nonuniform work function distribution on a surface. In this work, we let the $x$-axis run along the surface of the cathode, perpendicular to the patch edges, and the $y$-axis run along the cathode-anode direction, i.e., the direction of electron emission away from the surface. The cathode is set to be at $y = 0$, while the anode at $y = d$. Mathematically, the work function as a function of surface position is:

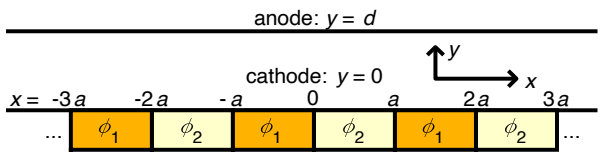

Fig. 1. Model heterogeneous emission surface characterized as a "zebra crossing" spatial distribution of work function in a laterally infinite parallel diode.

$$\phi(x) = \begin{cases} \phi_1, & (2N-1)a < x < 2Na \\ \phi_2, & 2Na < x < (2N+1)a \end{cases} \quad (1)$$

where $\phi_1 > \phi_2 > 0$, $N = 0, \pm 1, \pm 2, \ldots$ is any integer.

### B. Quantifying the shape of Miram curve knee

We define the TL-FSCL intersection temperature $T_i$ as the intersection temperature of the separate extrapolations of temperature-limited (TL) $J_{TL}$ and full-space-charge-limited (FSCL) $J_{FSCL}$ curves (Fig. 2):

$$J_{TL}(T_i) = J_{FSCL}(T_i) \quad (2)$$

As Fig. 2 shows, much of a Miram curve knee may be observed at temperatures greater than the TL-FSCL intersection temperature $T_i$. (More discussions in Section III)

In this work, we use an area-averaged application of the Richardson-Laue-Dushman (RLD) equation to estimate the TL extrapolation. For the zebra crossing work function distribution with equal widths for the two work function stripes (Fig. 1), the TL extrapolation equation is

$$J_{TL} = \frac{1}{2}\left[AT^2 \exp\left(-\frac{\phi_1}{kT}\right) + AT^2 \exp\left(-\frac{\phi_2}{kT}\right)\right] \quad (3)$$

where $T$ is the temperature, $k$ is the Boltzmann constant, and the Richardson constant $A = 4\pi m e k^2 / h^3$ where $m$ is the electron mass, $e$ is the elementary charge, and $h$ is the Planck constant.

The FSCL extrapolation can be estimated using the Child-Langmuir law with finite temperature correction (CLT). The equation at the TL-FSCL intersection temperature $T_i$ for a uniform cathode with a single cathode work function value $\phi_K$ is:[8], [17]

$$J_{CLT} = \frac{4\epsilon_0}{9}\sqrt{\frac{2e}{m}}\frac{V_{AK}^{\frac{3}{2}}}{d^2}\frac{9}{8\sqrt{\pi}}\eta^{-\frac{3}{2}}\left(\int_0^\eta \frac{d\eta}{\sqrt{\operatorname{erfcx}\sqrt{\eta}-1+2\sqrt{\eta/\pi}}}\right)^2 \quad (4)$$

where $\epsilon_0$ is the vacuum permittivity, $V_{AK} = V_{applied} - \phi_A/e + \phi_K/e$ is the difference of the electric potential of the vacuum level between the anode surface and cathode surface, $V_{applied}$ is the applied voltage between the anode and the cathode as measured in experiments, $\phi_A$ is the anode work function, $\phi_K$ is the cathode work function, $\eta = eV_{AK}/(kT)$, and erfcx is the scaled complementary error function.

For a zebra crossing cathode, we average $J_{CLT}$ over the cathode surface to estimate the FSCL extrapolation $J_{FSCL}$:

$$J_{FSCL} = \frac{1}{2}\left[J_{CLT}|_{\phi_K=\phi_1} + J_{CLT}|_{\phi_K=\phi_2}\right] \quad (5)$$

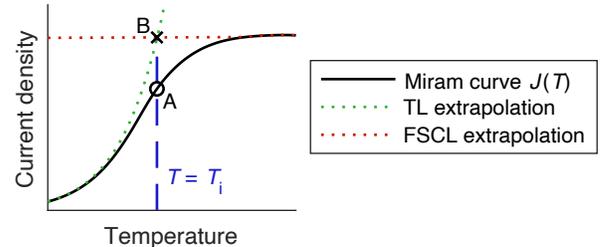

Fig. 2. Schematic plot illustrating the definition of the TL-FSCL intersection temperature $T_i$ (dashed blue line) and the normalized intersection emission parameter $\alpha$. In this plot, $\alpha = J(A)/J(B)$. The solid black curve is the Miram curve $J(T)$. The dotted red curve is the full-space-charge-limited (FSCL) extrapolation $J_{FSCL}$. The dotted green curve is the temperature-limited (TL) extrapolation $J_{TL}$.

To quantify the impact of the space charge and patch field effects on the emission current density at the TL-FSCL intersection temperature $T_i$, we define and use the intersection-temperature emission normalized by the FSCL current parameter (Fig. 2):

$$\alpha = \frac{J(T_i)}{J_{TL}(T_i)} = \frac{J(T_i)}{J_{FSCL}(T_i)} = \frac{J(A)}{J(B)} \quad (6)$$

A high $\alpha$ value represents good cathode performance and a Miram curve with a sharp knee at a low temperature and a flat saturated emission current. (More discussions in Section III)

Our previous study on nonuniform emission[7] indicated that space charge and patch fields have a significant impact on the shape of the Miram curve knee. Considering the definition of the normalized emission parameter (Eq. 6), we here develop complementary, limiting-case analytic models to separately study the following effects: a study of the space charge effect to analyze $J(T)/J_{FSCL}(T)$ and a study of the patch field effect to analyze $J(T)/J_{TL}(T)$.

### C. Space charge effect

Previous studies on 2-D and 3-D space charge[5], [13] illustrate its effects on the emission current. Seminal work from Lau developed a simple theory for the 2-D Child-Langmuir law.[18] We generalize this 2-D Child-Langmuir law to a cathode with nonuniform emission. In this generalized theory, we make some assumptions as detailed below.

<u>Assumption 1</u> (same as in Lau's work[18]): The $y$-axis component of the electric field at $(x_0, 0)$ is

$$E_y(x,0) = -\frac{\bar{V}_{AK}}{d} + G \int_0^d dy' \int_{-\infty}^{+\infty} dx' \frac{\rho(x',y')y'}{2\pi\epsilon_0[(x'-x)^2 + y'^2]} \quad (7)$$

where $G$ is a constant multiplication factor to account for the contributions due to the image charge. For a zebra crossing model (Fig. 1), the averaged electric potential difference between the anode and cathode surfaces is $\bar{V}_{AK} = V_{applied} - \phi_A/e + (\phi_1 + \phi_2)/(2e)$.

To solve for the value of $G$, we use the 1-D Child-Langmuir (CL) theory at zero temperature for a uniform cathode. In the space-charge-limited region, $E_y(x,0) = 0$, and

$$\rho(x,y) = J_{CL}\left(\frac{d}{y}\right)^{2/3}\sqrt{\frac{m}{2e\bar{V}_{AK}}} \quad (8)$$

where the Child-Langmuir (CL) current density

$$J_{CL} = \sqrt{\frac{2e}{m}}\frac{4\epsilon_0 \bar{V}_{AK}^{3/2}}{9d^2} \quad (9)$$

Solving Eq. 7 – 9, we get $G = 3/2$.

<u>Assumption 2</u>: The charge density has the same functional form as the Child-Langmuir law (Eq. 8) but replacing $J_{CL}$ with $J(x)$:

$$\rho(x,y) = J(x)\left(\frac{d}{y}\right)^{2/3}\sqrt{\frac{m}{2e\bar{V}_{AK}}} \quad (10)$$

We make this assumption because we are interested in the knee behavior where the emission current is close to $J_{FSCL}$ and the charge density $\rho(x,y)$ is close to the Child-Langmuir law case.

<u>Assumption 3</u>: For the zebra crossing case, we assume:
(1) The local emission current density over the $\phi_1$ patch is spatially uniform: $J(x) = J_1$, for $2Na - a < x < 2Na$ where $N$ is any integer.

(2) The $\phi_1$ patch emits in the temperature-limited (TL) region when the electric field at the center of the $\phi_1$ patch surface is negative, $E_{1y} = E_y(2Na - a/2, 0) < 0$.

(3) The space-charge-limited (SCL) emission current density is also spatially uniform over the $\phi_1$ patch, denoted as $J_{1SCL}$. which is determined by the equation $E_{1y} = 0$.

(4) The emission current density of the $\phi_1$ patch is $J_1 = \min\{J_{1TL}, J_{1SCL}\}$.

(5) Similar assumptions (1)-(4) also apply for the $\phi_2$ patch.

<u>Assumption 4</u>: As $\phi_1 > \phi_2 > 0$, the TL current densities have a relation $J_{2TL} > J_{1TL} > 0$, so the high work function patch ($\phi_1$) will reach SCL at a higher temperature than the low work function patch ($\phi_2$). It is assumed that the SCL current density for the high work function patch is equal to the 1-D Child-Langmuir theory $J_{1SCL} = J_{CL}$ while $J_{2SCL}$ may be larger than $J_{CL}$ at certain temperatures. This assumption is in agreement with the findings in previous studies[5], [13].

Under these assumptions, the electric field at the center of the low work function patch ($\phi_2$) is:

$$E_{2y} = \frac{V_{AK}}{d}\left(\frac{J_1}{J_{CL}} + \frac{J_2 - J_1}{J_{CL}}f - 1\right) \quad (11)$$

where $f$ is a function of only $a/d$:

$$f\left(\frac{a}{d}\right) = \frac{1}{3\pi}\int_0^1 d\bar{y} \sum_{N=-\infty}^{\infty} \int_{(2N-1/2)a/d}^{(2N+1/2)a/d} \frac{\bar{y}^{1/3}}{\bar{x}^2 + \bar{y}^2} d\bar{x} \quad (12)$$

where the symbols with bars are normalized coordinates $\bar{x} = x/d$, $\bar{y} = y/d$.

Letting $E_{2y} = 0$, we can solve for the space-charge-limited (SCL) current density of the $\phi_2$ patch: $J_{2SCL} = J_1 + (J_{CL} - J_1)/f$. Therefore, we can predict the average emission current density for the zebra crossing cathode:

(1) Temperature-limited (TL) region $J_{1TL} < J_{2TL} < J_{2SCL}$: Neither the $\phi_1$ patch nor the $\phi_2$ patch reaches SCL. The average current density of the cathode is $J = (J_{1TL} + J_{2TL})/2$.

(2) Full-space-charge-limited (FSCL) region $J_{2TL} > J_{1TL} > J_{1SCL} = J_{CL}$: Both the $\phi_1$ and $\phi_2$ patches are SCL, so $J = (J_{1SCL} + J_{2SCL})/2 = J_{CL}$.

(3) TL-FSCL transition region (i.e., the location of the Miram curve knee) $J_{2TL} > J_{2SCL}$ but $J_{1TL} < J_{1SCL}$: The low work function patch ($\phi_2$) is space-charge-limited (SCL) while the high work function patch ($\phi_1$) is still not SCL, so $J = (J_{1TL} + J_{2SCL})/2 = J_{CL} - (J_{CL} - J_{1TL})(2f - 1)/(2f)$.

In particular, at the TL-FSCL intersection temperature $T_i$, we assume that the temperature satisfies the condition:

$$J_{CL} = J_{TL} = \frac{1}{2}\left[AT_i^2 \exp\left(-\frac{\phi_1}{kT_i}\right) + AT_i^2 \exp\left(-\frac{\phi_2}{kT_i}\right)\right] \quad (13)$$

The normalized intersection emission restricted by the space charge effect is

$$\alpha_{SC} = \frac{J(T_i)}{J_{CL}} = 1 - \frac{2f - 1}{2f}\tanh\frac{\phi_1 - \phi_2}{2kT_i} \quad (14)$$

which only depends on two parameters: (1) the normalized work function difference $(\phi_1 - \phi_2)/(kT_i)$ and (2) the patch-diode size ratio $a/d$ (cf. Eq. (12): $f$ only depends on $a/d$).

### D. Patch field effect

We assume the additional electric potential due to the patch field effect satisfies $\nabla^2 V_{\text{PF}}(x,y) = 0$, with the boundary condition for the cathode surface $V_{\text{PF}}(x,0) = -\phi(x)/e$. We also assume that the anode-cathode distance is much larger than the patch width $d \gg a$, so use $E_y(x,\infty) = 0$ as the boundary condition for the anode side.

In this section, we adopt an approximate linear potential assumption, i.e., $\partial V_{\text{linear}}/\partial y = -E_0$, where $E_0 \leq 0$ is a uniform electric field, to estimate the effects of the applied voltage and the space charge on the electric potential. Under these assumptions, the electric potential is $V = V_{\text{PF}} + V_{\text{linear}} = V_{\text{PF}} - E_0 y$.

Considering the periodicity and the symmetry of this problem, here we only show the results in $-a/2 < x < a/2$:

$$V(x,y) = -\frac{\phi_1 + \phi_2}{2e} + \frac{\phi_1 - \phi_2}{e} p(x,y,\beta_E) \quad (15)$$

where the field ratio

$$\beta_E = \frac{E_{\text{patch}}}{-E_0} = \frac{\phi_1 - \phi_2}{-eaE_0} \quad (16)$$

and the parameter

$$p(x,y,\beta_E) = \frac{\breve{y}}{\pi \beta_E} + \frac{i}{\pi}\left(\text{arctanh}\, e^{-i\breve{x}-\breve{y}} - \text{arctanh}\, e^{i\breve{x}-\breve{y}}\right) \quad (17)$$

where the symbols with breves are normalized coordinates $\breve{x} = \pi x/a$, $\breve{y} = \pi y/a$, and i is the imaginary unit.

The voltage minimum is $V_m(x) = \min_y V(x,y) = V(x, y_m(x))$, where the location of the voltage minimum $y_m(x)$ satisfies

$$\cosh \breve{y}_m(x) = (\beta_E \sin \breve{x})/2 + \sqrt{1 + \left[\left(\frac{\beta_E}{2}\right)^2 - 1\right] \sin^2 \breve{x}} \quad (18)$$

for $0 < \breve{x} < \min\{\pi/2, \arcsin \beta_E\}$, and $y_m(x) = 0$ for elsewhere.

For a given cathode surface location $x$, the location of the voltage minimum $y_m(x)$ only depends on the field ratio $\beta_E$. Therefore, for a given $x$, the parameter $p(x, y_m(x), \beta_E)$ in Eq. 17 also only depends on $\beta_E$.

The local emission current density follows the RLD equation $J(x) = AT^2 \exp[eV_m(x)/(kT)]$. The cathode-averaged emission current density $J$ can be obtained by averaging $J(x)$ over the cathode surface. The ratio of the average emission current density $J$ to the TL extrapolation $J_{\text{TL}}$ (Eq. 3) is:

$$\frac{J}{J_{\text{TL}}} = \frac{\frac{1}{\pi}\int_{-\frac{\pi}{2}}^{\frac{\pi}{2}} AT^2 \exp\left[\frac{eV_m(x)}{kT}\right] d\breve{x}}{\frac{1}{2}\left[AT^2 \exp\left(-\frac{\phi_1}{kT}\right) + AT^2 \exp\left(-\frac{\phi_2}{kT}\right)\right]}$$

$$= \frac{\frac{1}{\pi}\int_{-\frac{\pi}{2}}^{\frac{\pi}{2}} \exp\left[\frac{\phi_1 - \phi_2}{kT} p(x, y_m(x), \beta_E)\right] d\breve{x}}{\cosh \frac{\phi_1 - \phi_2}{2kT}} \quad (19)$$

which only depends on the normalized work function difference $(\phi_1 - \phi_2)/(kT)$ and the field ratio $\beta_E$.

In the low temperature limit where the space charge effect is negligible, $-E_0 = E_{\text{AK}} = V_{\text{AK}}/d$. To estimate the effect of the space charge on the $E_0$ value, we made the similar assumption as Eq. 10 but replacing $J(x)$ with its average value $J$:

$$\rho(x,y) = J\left(\frac{d}{y}\right)^{2/3} \sqrt{\frac{m}{2eV_{\text{AK}}}} \quad (20)$$

Substituting Eq. 20 into Eq. 7, we get that the electric field $E_0$ (or $E_y(x,0)$ in Eq. 7) due to the applied voltage and the space charge satisfies:

$$1 = \frac{-E_0}{E_{\text{AK}}} + \frac{J}{J_{\text{FSCL}}} = \frac{E_{\text{patch}}}{E_{\text{AK}}} \frac{1}{\beta_E} + \frac{J}{J_{\text{TL}}} \frac{J_{\text{TL}}}{J_{\text{FSCL}}} \quad (21)$$

where $\beta_E = E_{\text{patch}}/(-E_0)$.

When the value of $J_{\text{TL}}/J_{\text{FSCL}}$ is given, we eliminate the variable $\beta_E$ by solving the system of Eqs. 19 and 21, so $J/J_{\text{TL}}$ only depends on $(\phi_1 - \phi_2)/(kT)$ and $E_{\text{patch}}/E_{\text{AK}}$.

In particular, at the TL-FSCL intersection temperature $T_i$, $J_{\text{TL}}/J_{\text{FSCL}} = 1$ and the normalized intersection emission restricted by the patch field effect $\alpha_{\text{PF}} = J(T_i)/J_{\text{TL}}(T_i)$ also only depends on two parameters: (1) the normalized work function difference $(\phi_1 - \phi_2)/(kT_i)$ and (2) the patch-diode field ratio $E_{\text{patch}}/E_{\text{AK}} = [(\phi_1 - \phi_2)/(ea)]/(V_{\text{AK}}/d)$.

## III. RESULTS AND DISCUSSION

We develop complementary, limiting-case analytic theory models to separately estimate the effects of space charge and patch fields on the shape of the Miram curve. From these models, we reveal there are three main physical parameters significantly affecting the shape of the Miram curve knee: (1) the normalized work function difference $(\phi_1 - \phi_2)/(kT_i)$, (2) the patch-diode size ratio $a/d$, and (3) the patch-diode field ratio $E_{\text{patch}}/E_{\text{AK}} = [(\phi_1 - \phi_2)/(ea)]/(V_{\text{AK}}/d)$.

### A. Shape of Miram curve knee

Fig. 3 shows the Miram curves predicted by our analytic models in different cases, compared with the simulation results using the nonuniform emission model[7] and their TL and FSCL extrapolations. The values of the three main physical parameters, the TL-FSCL intersection temperature $T_i$ calculated using Eq. 2, and the normalized intersection emission $\alpha$ values defined in Eq. 6 for each case in Fig. 3 are listed in Table I.

In Figs. 3a and 3b, the patch fields have more significant impact than the space charge in restricting the emission current. The Miram curves predicted by our patch field limiting case theoretical model (dashed magenta curves) are consistent with simulation results (solid black curve)[7].

When $E_{\text{patch}}/E_{\text{AK}} \gg 1$ (Fig. 3a), there is a significant discrepancy between the simulated Miram curve (solid black curve) and its TL extrapolation (dotted green curve, which overlaps with the dashed sky-blue space charge limiting case model curve at low temperatures). In this case, the patch field effect significantly lowers the emission current in TL region, resulting in a knee temperature much higher than the intersection temperature $T_i$ of the TL and FSCL extrapolations (Points A and B). The normalized intersection emission $\alpha$ value is very low.

When $E_{\text{patch}}/E_{\text{AK}} \approx 1$ (Fig. 3b), the patch field effect moderately lowers the emission current in TL region and leads to a very rounded knee. In this case, the knee temperature is slightly higher than $T_i$, and $\alpha$ is moderately low.

Fig. 3c is an example where the impacts of both effects are comparable and not negligible. The $\alpha$ value is high (close to 1).

In Fig. 3d, the space charge effect restricts the emission current, causing the saturated emission current density to be not flat. In this case, the simulated Miram curve is step-shaped (solid black curve), consistent with the results predicted by our limit case theoretical model of the space charge effect (dashed sky-blue curve), and $\alpha$ is again moderately low.

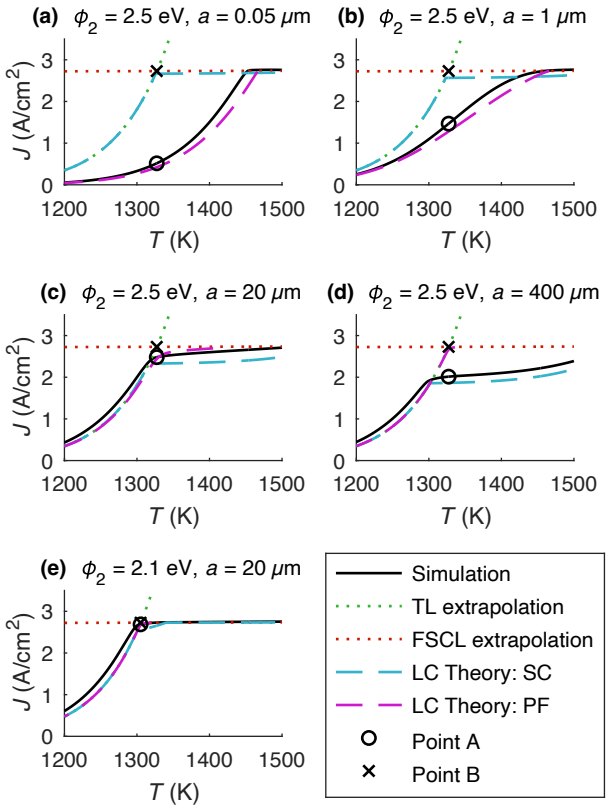

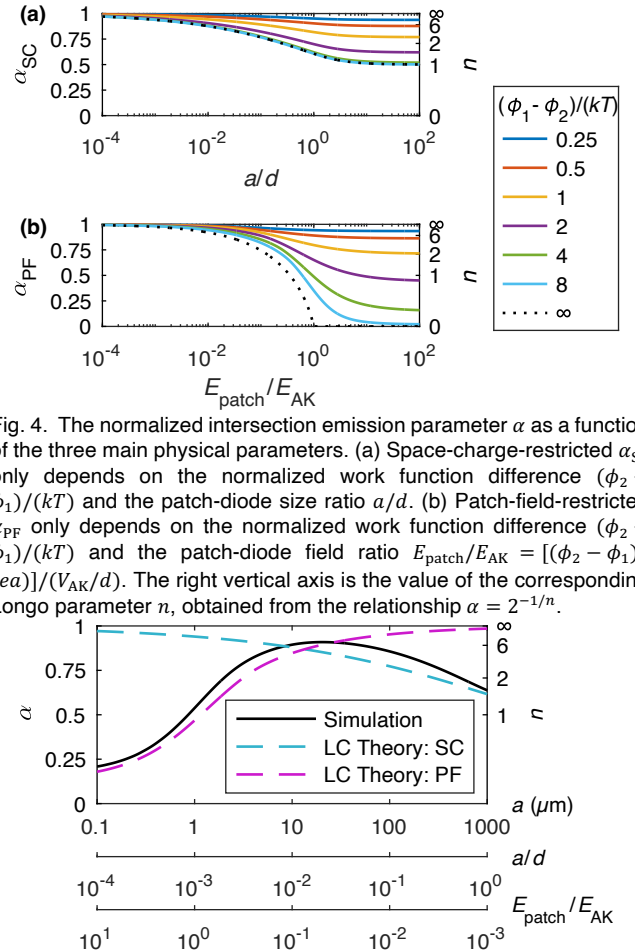

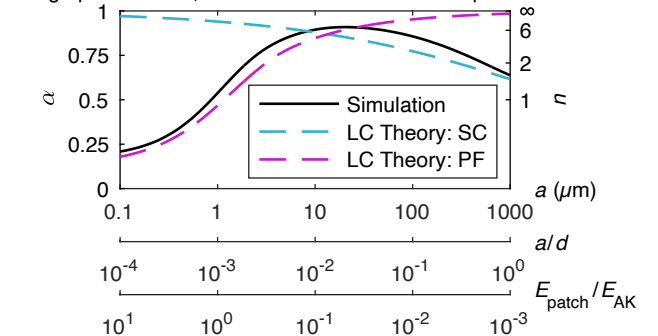

Fig. 3. Predicted Miram curves by limiting-case (LC) theories (dashed sky-blue curve for space charge limiting case, dashed magenta curve for patch field limiting case) compared with the simulation results of nonuniform emission model (solid black curve), from a zebra crossing work function distribution, with $\phi_1 = 2$ eV, $d = 1$ mm, $V = 500$ V. The dotted green and red curves are the TL and FSCL extrapolation, respectively. The "o" and "x" symbols indicate the TL-FSCL intersection temperature $T_i$ calculated using Eq. 2. "o" is the Point A in Fig. 2, while "x" is Point B. Different subfigures have different values of $\phi_2$ and the stripe width $a$, as shown in the title of each subfigure.

TABLE I
VALUES OF THE PARAMETERS IN THE CASES IN FIG. 3

| Sub-figure | $\dfrac{\phi_2 - \phi_1}{kT}$ | $\dfrac{a}{d}$ | $\dfrac{E_{\text{patch}}}{E_{\text{AK}}}$ | $T_i$ (K) | $\alpha$ |
|---|---|---|---|---|---|
| (a) | 4.37 | 0.00005 | 20 | 1327 | 0.1890 |
| (b) | 4.37 | 0.001 | 1 | 1327 | 0.5380 |
| (c) | 4.37 | 0.02 | 0.05 | 1327 | 0.9090 |
| (d) | 4.37 | 0.4 | 0.0025 | 1327 | 0.7378 |
| (e) | 0.89 | 0.02 | 0.01 | 1305 | 0.9853 |

The normalized work function difference $(\phi_1 - \phi_2)/(kT_i)$ governs the impact of both the patch field and space charge effects. Decreasing this parameter is one of the methods to lessen the restriction on the emission current due to both patch fields and space charge effects, resulting in a sharp knee at low temperature and a flat saturated emission current density (Fig. 3e). In this case, the $\alpha$ value is very high, close to 1.

### B. Physical factors impacting normalized emission at TL-FSCL intersection temperature

As Fig. 3 shows, a large value of the normalized emission value $\alpha$ at TL-FSCL intersection temperature represents good cathode performance, indicating that neither the space charge nor the patch field effects significantly restrict the knee-temperature emission current and that the Miram curve has a sharp knee at a temperature close to $T_i$ with a flat saturated emission current. A small $\alpha$ value represents that the space charge (local to the cathode) and/or patch field effects significantly restrict the knee-temperature emission current, resulting in a rounded knee, an increased knee temperature, and/or an inclined saturated emission current.

To further illustrate the quantitative effects of the physical factors to the Miram curve knee, in Fig. 4 we plot the predicted $\alpha$ value as a function of the three main physical parameters, using our theoretical studies in Section II. The lower these three parameters are, the higher the normalized intersection emission parameter $\alpha$ is, implying a sharper knee at a lower temperature with a flatter saturated emission current and a better cathode performance.

Figs. 3a – 3d show the results where the values of work function, the anode-cathode distance, and the anode-cathode voltage are fixed but the patch size $a$ is variable. Fig. 5 illustrates the relationship between the normalized intersection emission parameter $\alpha$ and patch size $a$. In the cases of a small

patch size, $a/d$ is small but $E_\text{patch}/E_\text{AK}$ is large, so the emission tends to be patch-field-restricted (Figs. 3a and 3b). In the cases of a large patch size, $a/d$ is large but $E_\text{patch}/E_\text{AK}$ is small, so the emission tends to be space-charge-restricted (Fig. 3d). For the illustrative parameters of Fig. 3, the boundary between the patch-field-restricted and the space-charge-restricted regimes approximately locates at $a \approx 20$ μm (Fig. 3c). This result indicates that for a zebra crossing model, when $\phi_1$, $\phi_2$, $d$, and $V$ are fixed, there exists an optimal patch size $a$ which leads to a highest normalized intersection emission $\alpha$. This indicates that for a given cathode material (work function values fixed) and a given diode fixture ($d$ and $V$ fixed), there may exist an optimal effective emission patch size so that the cathode can have the best emission performance as it relates to the shape of Miram curve knee.

### C. Relationship to Longo-Vaughan equation

The Longo-Vaughan equation[19] $J_\text{TL}^{-n} + J_\text{FSCL}^{-n} = J^{-n}$ is a commonly used empirical equation to describe the smooth rounded knee in Miram curves. Solving the system of the Longo-Vaughan equation and Eq. 2, we get a relationship between the Longo-Vaughan parameter $n$ and the normalized knee emission parameter $\alpha$, i.e., $\alpha = 2^{-1/n}$. Such a relation is plotted in the vertical axis on the right in Figs. 4 and 5.

Vaughan[19] pointed out that well-designed and well-made electron guns have $n$ values in the range of 6 to 10, which corresponds to $\alpha$ values of 0.89 to 0.93, while diodes with so-called "patchy emission", ascribed to uneven heating or other defects, have $n$ in the 2 to 5 range, corresponding to $\alpha$ values of 0.71 to 0.87.

## IV. CONCLUSION

In this work, we used an equal-width periodic striped (zebra crossing) work function spatial distribution to study nonuniform thermionic emission. We used a normalized emission parameter $\alpha$ at the TL-FSCL intersection temperature to quantify the shape of the Miram curve knees. We developed complementary, limiting-case analytic models to separately estimate the effects of space charge and patch fields, and found that there are three main physical parameters which significantly affect the shape of the knee: (1) the normalized work function difference $(\phi_2 - \phi_1)/(kT)$, (2) the patch-diode size ratio $a/d$, and (3) the patch-diode field ratio $E_\text{patch}/E_\text{AK}$. The lower these three parameters are, the higher the $\alpha$ value is, implying a sharper knee at a lower temperature with a flatter saturated emission current, and a better cathode performance.

The physical knowledge revealed in this work directly and quantitatively connects the patch size, work function values, anode-cathode voltage, and anode-cathode gap distance to the shape of the Miram curve knee, providing new understanding and a guide to the design of thermionic cathodes used as electron sources in vacuum electronic devices (VEDs).

## REFERENCES

[1] R. J. Barker, N. C. Luhmann, J. H. Booske, and G. S. Nusinovich, *Modern microwave and millimeter-wave power electronics*. 2005.

[2] J. H. Booske, "Plasma physics and related challenges of millimeter-wave-to-terahertz and high power microwave generation," *Phys. Plasmas*, vol. 15, no. 5, p. 055502, May 2008, doi: 10.1063/1.2838240.

[3] J.-Y. Gao, Y.-F. Yang, X.-K. Zhang, S.-L. Li, P. Hu, and J.-S. Wang, "A review on recent progress of thermionic cathode," *Tungsten*, no. 0123456789, pp. 1–15, Sep. 2020, doi: 10.1007/s42864-020-00059-1.

[4] M. J. Cattelino, G. V. Miram, and W. R. Ayers, "A diagnostic technique for evaluation of cathode emission performance and defects in vehicle assembly," in *1982 International Electron Devices Meeting*, 1982, pp. 36–39, doi: 10.1109/IEDM.1982.190205.

[5] D. Chernin, Y. Y. Lau, J. J. Petillo, S. Ovtchinnikov, D. Chen, A. Jassem, R. Jacobs, D. Morgan, and J. H. Booske, "Effect of Nonuniform Emission on Miram Curves," *IEEE Trans. Plasma Sci.*, vol. 48, no. 1, pp. 146–155, Jan. 2020, doi: 10.1109/TPS.2019.2959755.

[6] V. Vlahos, "Private communication." Private communication, 2021.

[7] D. Chen, R. Jacobs, D. Morgan, and J. Booske, "Impact of Nonuniform Thermionic Emission on the Transition Behavior Between Temperature-and Space-Charge-Limited Emission," *IEEE Trans. Electron Devices*, vol. 68, no. 7, pp. 3576–3581, Jul. 2021, doi: 10.1109/TED.2021.3079876.

[8] D. Chen, R. Jacobs, J. Petillo, V. Vlahos, K. L. Jensen, D. Morgan, and J. Booske, "Physics-based Model for Nonuniform Thermionic Electron Emission from Polycrystalline Cathodes," pp. 1–29, Nov. 2021, [Online]. Available: http://arxiv.org/abs/2112.02136.

[9] K. T. Compton and I. Langmuir, "Electrical Discharges in Gases. Part I. Survey of Fundamental Processes," *Rev. Mod. Phys.*, vol. 2, no. 2, pp. 123–242, Apr. 1930, doi: 10.1103/RevModPhys.2.123.

[10] J. A. Becker, "Thermionic Electron Emission and Adsorption Part I. Thermionic Emission," *Rev. Mod. Phys.*, vol. 7, no. 2, pp. 95–128, Apr. 1935, doi: 10.1103/RevModPhys.7.95.

[11] W. B. Nottingham, "Thermionic Emission from Tungsten and Thoriated Tungsten Filaments," *Phys. Rev.*, vol. 49, no. 1, pp. 78–97, Jan. 1936, doi: 10.1103/PhysRev.49.78.

[12] C. Herring and M. H. Nichols, "Thermionic Emission," *Rev. Mod. Phys.*, vol. 21, no. 2, pp. 185–270, Apr. 1949, doi: 10.1103/RevModPhys.21.185.

[13] A. Jassem, D. Chernin, J. J. Petillo, Y. Y. Lau, A. Jensen, and S. Ovtchinnikov, "Analysis of Anode Current From a Thermionic Cathode With a 2-D Work Function Distribution," *IEEE Trans. Plasma Sci.*, pp. 1–7, 2021, doi: 10.1109/TPS.2020.3048097.

[14] A. Sitek, K. Torfason, A. Manolescu, and Á. Valfells, "Space-Charge Effects in the Field-Assisted Thermionic Emission from Nonuniform Cathodes," *Phys. Rev. Appl.*, vol. 15, no. 1, p. 014040, Jan. 2021, doi: 10.1103/PhysRevApplied.15.014040.

[15] L. K. Hansen, "Anomalous Schottky Effect," *J. Appl. Phys.*, vol. 37, no. 12, pp. 4498–4502, Nov. 1966, doi: 10.1063/1.1708068.

[16] T. Schultz, T. Lenz, N. Kotadiya, G. Heimel, G. Glasser, R. Berger, P. W. M. Blom, P. Amsalem, D. M. de Leeuw, and N. Koch, "Reliable Work Function Determination of Multicomponent Surfaces and Interfaces: The Role of Electrostatic Potentials in Ultraviolet Photoelectron Spectroscopy," *Adv. Mater. Interfaces*, vol. 4, no. 19, p. 1700324, Oct. 2017, doi: 10.1002/admi.201700324.

[17] I. Langmuir, "The Effect of Space Charge and Initial Velocities on the Potential Distribution and Thermionic Current between Parallel Plane Electrodes," *Phys. Rev.*, vol. 21, no. 4, pp. 419–435, Apr. 1923, doi: 10.1103/PhysRev.21.419.

[18] Y. Y. Lau, "Simple Theory for the Two-Dimensional Child-Langmuir Law," *Phys. Rev. Lett.*, vol. 87, no. 27, pp. 278301-278301–3, 2001, doi: 10.1103/PhysRevLett.87.278301.

[19] R. Vaughan, "A synthesis of the Longo and Eng cathode emission models," *IEEE Trans. Electron Devices*, vol. 33, no. 11, pp. 1925–1927, Nov. 1986, doi: 10.1109/T-ED.1986.22844.